\documentclass[DIV=calc, paper=a4, fontsize=11pt, twocolumn]{scrartcl}	 

\usepackage[english]{babel} 
\usepackage[protrusion=true,expansion=true]{microtype} 

\usepackage[sc]{mathpazo} 
\usepackage[T1]{fontenc} 
\usepackage{amsmath,amsfonts,amsthm} 
\usepackage[svgnames]{xcolor} 
\usepackage[hang, small,labelfont=bf,up,textfont=it,up]{caption} 
\usepackage{booktabs} 
\usepackage{fix-cm}	 
\usepackage{physics}

\usepackage[absolute]{textpos} 

\usepackage{graphicx}
\graphicspath{{figures/}}
\usepackage{hyperref}

\usepackage{scrlayer}
\usepackage{lastpage} 
\usepackage[none]{hyphenat}

\usepackage[T1]{fontenc}
\usepackage{lmodern}

\usepackage{titling} 

\usepackage{titling} 

\newcommand{\HorRule}{\color{DodgerBlue} \rule{\linewidth}{1pt}} 

\pretitle{\vspace{-30pt} \begin{center} \HorRule \fontsize{30}{30} \usefont{OT1}{phv}{b}{n} \color{Blue} \selectfont} 

\title{Leveraging modular values in quantum algorithms: \\the Deutsch-Jozsa} 

\posttitle{\par\end{center}\vskip 0.5em} 

\preauthor{\begin{center}\large \lineskip 0.5em \usefont{OT1}{phv}{b}{sl} \color{DodgerBlue}} 

\author{Lorena Ballesteros Ferraz$~^{1,2,3,4}$, Timoteo Carletti$~^{2,5}$, and Yves Caudano$~^{1,2,3}$} 

\postauthor{\vspace{10pt}\footnotesize \usefont{OT1}{phv}{m}{sl} \color{Black} 
\\$^{1}$naXys, Namur Institute for Complex Systems, University of Namur, Belgium \\
$^{2}$NISM, Namur Institute of Structured Matter, University of Namur, Belgium\\ 
$^{3}$Research Unit Lasers and Spectroscopies (UR-LLS), Department of Physics, University of Namur, Belgium\\
$^{4}$Present adress: Université de Lorraine, CNRS, LPCT, F-54000 Nancy, France\\
$^{5}$Department of Mathematics, University of Namur, Belgium
\date{} 

\par\end{center}\HorRule} 

\begin{document}

\twocolumn[
  \begin{@twocolumnfalse}
    \maketitle
        \vspace{-40pt}

 \begin{abstract}
   \textbf{We present a novel approach to quantum algorithms, by taking advantage of modular values, i.e., complex and unbounded quantities resulting from specific post-selected measurement scenarios. Our focus is on the problem of ascertaining whether a given function acting on a set of binary values is constant (uniformly yielding outputs of either all $0$ or all $1$), or balanced (a situation wherein half of the outputs are $0$ and the other half are $1$). Such problem can be solved by relying on the Deutsch-Jozsa algorithm. The proposed method, relying on the use of modular values, provides a high number of degrees of freedom for optimizing the new algorithm inspired from the Deutsch-Jozsa one. In particular,  we  explore meticulously the choices of the pre- and post-selected states. We eventually test the novel theoretical algorithm on a quantum computing platform. While the outcomes are currently not on par with the conventional approach, they nevertheless shed light on potential for future improvements, especially with less-optimized algorithms. We are thus confidend that the proposed proof of concept could prove its validity in bridging quantum algorithms and modular values research fields.}
    \end{abstract}
  \end{@twocolumnfalse}
    \vspace{20pt}
  ]



\section*{Introduction}
Numerous quantum algorithms have demonstrated theoretically superior efficiency compared to their presently known classical counterparts, with historical examples including database search \cite{grover1996fast} and integer factorization \cite{shor1994algorithms}. Among the earliest instances of this superiority are the Deutsch algorithm \cite{deutsch1992rapid} and its generalization, the Deutsch-Jozsa algorithm \cite{cleve1998quantum}. The Deutsh-Jozsa algorithm addresses the verification of whether a binary function acting on $n$ classical bits, $f:\{0,1\}^n\rightarrow\{0,1\}$, is balanced or constant. The function is constant if it has the same output for all possible values of the $n$ input bits, while it is balanced if it is equal to $0$ for exactly half of the possible values of the input bits (and $1$ for the other half). Remarkably, the quantum approach accomplishes this task in a single step, while the best classical algorithm necessitates $2^n-1$ repetitions in the worst-case scenario.\\
A weak value $A_w=\frac{\bra{\psi_f}\hat{A}\ket{\psi_i}}{\bra{\psi_f}\ket{\psi_i}}$ is an unbounded complex number involving a Hermitian operator $\hat{A}$, as well as an initial state $\ket{\psi_i}$ and a final state $\ket{\psi_f}$, respectively called pre- and post-selected states in this context. The concept of weak values of observables arises from a quantum process, called weak measurement, that involves pre- and post-selection, as introduced in the work of Aharonov, Albert, and Vaidman (AAV) in the 1980s \cite{aharonov1988result}. In this protocol, the whole quantum system is composed of an ancillary component acting as a measuring device and a primary system of interest. At the outset, this combined entity is described by the separable quantum state $\ket{\Psi}=\ket{\psi_i}\otimes\ket{\phi}$, where the pre-selected state $\ket{\psi_i}$ is the state in which the system of interest is set prior to the measurement, while $\ket{\phi}$ represents the initial state of the ancillary component. The protocol unfolds in four steps. Initially, the system's pre-selection occurs, involving the selection of the initial state $\ket{\psi_i}$ of the system of interest and the preparation of the ancillary state $\ket\phi$. Subsequently, an interaction occurs between the system and the ancillary component, which is mathematically modeled by the Hamiltonian operator $\hat{H}=\hbar\, g\!\left(t\right) \hat{A} \otimes \hat{p}$ inspired from the von Neumann protocol for quantum measurement. Here, the coupling parameter $g\!\left(t\right)$ determines the interaction coupling strength, which typically emerges as a small factor in the process, approximately given by $\gamma=\int_{t_i}^{t_f}g\!\left(t\right)dt \ll 1$ \cite{aharonov1988result}, where $t_f-t_i$ is the short interaction duration. The operator $\hat{p}$ represents the momentum operator of the measuring device. The observable $\hat{A}$ corresponds to the physical quantity probed by the protocol. The (weak) interaction entangles (slightly) the system of interest and the ancillary component. Finally, a post-selection stage follows: it involves a projective measurement and filtering applied to the system, throwing away cases in which the system of interest does not end up in the desired final state $\ket{\psi_f}$. Conditioned on a successful post-selection of $\ket{\psi_f}$, a measurement of the ancillary component is performed to extract information about the weak value, obtained from the shift of the average of a meter observable.  Typically, the ancilla's wavefunction in position representation is shifted by the real part of the weak value, while the ancilla's wavefunction in momentum representation is shifted by the imaginary part of the weak value.\\
The modular value, expressed as 
\begin{equation}
A_m=\frac{\bra{\psi_f}e^{-ik\hat{A}}\ket{\psi_i}}{\bra{\psi_f}\ket{\psi_i}},
\end{equation}
arises from a protocol similar to the one of weak measurements, in which pre- and post-selection is necessary. However, modular values are unique as they are associated with interactions of any strength because they  involve a qubit as the ancillary meter component \cite{kedem2010modular}, rendering a weak coupling between the system and ancilla unnecessary to evaluate  the final meter state in practice. 
After the process of pre-selection, interaction and post-selection, the final meter state takes the form, 
\begin{eqnarray}
&&\frac{1}{\sqrt{N}}\left(\alpha\ket{0}+\beta\frac{\bra{\psi_f}e^{-ikA}\ket{\psi_i}}{\bra{\psi_f}\ket{\psi_i}}\ket{1}\right) \\ \nonumber
&=&\frac{1}{\sqrt{N}}\left(\alpha\ket{0}+\beta A_m\ket{1}\right),
\end{eqnarray}
where the value $\frac{1}{\sqrt{N}}$ is a normalization factor \cite{kedem2010modular} and $\alpha\ket{0}+\beta\ket{1}$ is the meter initial state (prior to its interaction with the system of interest).\\
Modular values exhibit strong similarity to weak values under specific circumstances, particularly when $k$ is small, so that the exponential can be approximated by using a Taylor expansion. Remarkably, the modular value represents a weak value of a unitary observable, when the unitary operator is also Hermitian. In recent years, both quantum weak values and modular values have garnered considerable attention due to their wide-ranging applications in various fields, such as quantum foundations \cite{kocsis2011observing, matzkin2019weak, Rafiepoor2021, Cormann2017}, quantum tomography \cite{Lundeen2011, Pan2019, Turek2020}, and particularly in quantum metrology \cite{hosten2008observation, zhang2015precision, li2018chiral, Ho2019} and measurements \cite{Ho2016, Ho2017, Ho2018, Parks2018, Ogawa2019, Li2020} in general. The intriguing quantum phenomena associated with weak and modular values hold significant promise for advancing our understanding of quantum mechanics and finding practical uses in cutting-edge technologies.\\
This paper introduces a novel approach to implementing quantum algorithms by leveraging the unique characteristics of modular values, in practice by calculating the modular value of the oracle involved in a quantum algorithm. Our proposed procedure capitalizes on the degrees of freedom inherent to modular values, including the fact that they are complex numbers. However, it is worth noting that this method necessitates the use of one extra qubit to facilitate the readout of the result obtained from the quantum modular value. The model can be applied to various algorithms. Our work bridges two captivating quantum research fields, whose links remained elusive, urging us to explore connections between quantum algorithms and weak and modular values. While isolated studies have been conducted in this area \cite{pati2019super}, a comprehensive theoretical framework is yet to be established.\\
This paper is structured as follows. The initial section introduces a comprehensive approach to address quantum algorithms through modular values. Subsequently, these findings are applied to the distinct scenario associated with the Deutsch-Jozsa algorithm. Following this, we explore the diverse options for choosing adequately the pre- and post-selected states connected with the modular value. Furthermore, the outcomes of implementing this algorithm on an actual quantum computer---the IBM quantum computer---are showcased. Ultimately, the paper concludes by summarizing its contributions and outlining potential future directions.

\section{Exploring the implementation of quantum algorithm via modular values of oracles}
\begin{figure} [t!]
\centering 
\includegraphics[width=0.5\textwidth]{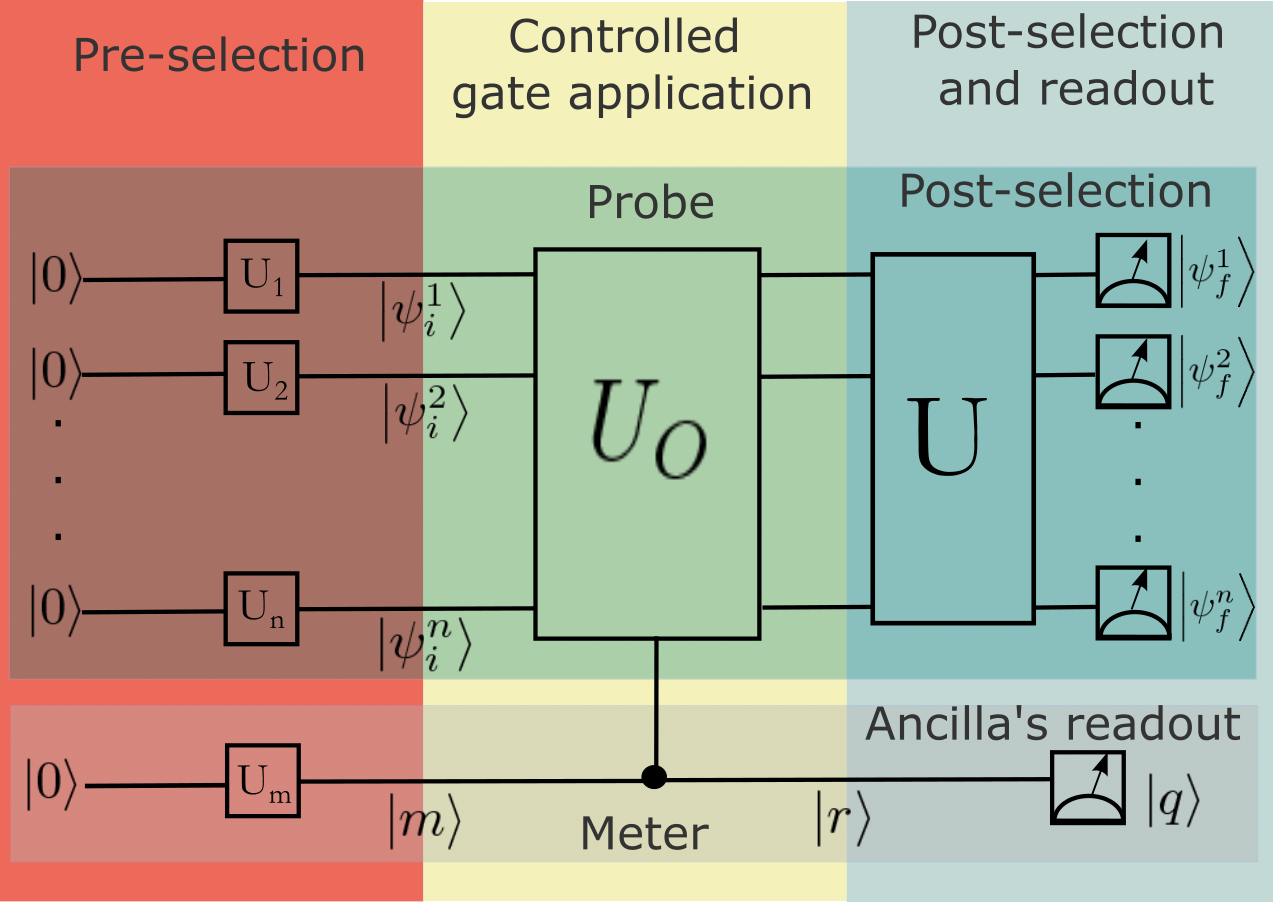}
  \caption{The proposed model consists of $n$ qubits plus a qubit meter that together form the full system. To prepare the system for the experiment, we first apply $n$ unitary operators to the first $n$ qubits to take them to the pre-selected state $\ket{\psi_i}$. We also use a unitary operator to set the qubit meter to its initial state. Next, we apply the controlled oracle to the system. To post-select, we apply additional unitary operators that could involve one or several qubits of the system. Finally, we measure the state of the full system in the computational basis, and the results of the joint measurement are then processed only on the condition of successful post-selection\label{fig:scheme_model}}
\end{figure}

In this section, we provide a detailed procedure for tackling quantum algorithms using modular values. Our method involves carrying out a measurement with pre- and post-selection to the oracle. Oracles play a critical role in constructing algorithms in quantum computers \cite{kashefi2002comparison, menon2021quantum}. Essentially, an oracle is a black box that realizes unknown operations to the person executing the experiment. Oracles are implemented by using unitary operations and are commonly employed in quantum algorithms, such as the Deutsch-Jozsa algorithm and Grover's algorithm \cite{grover1996fast, cleve1998quantum}.\\
As illustrated in Fig.~\ref{fig:scheme_model}, the complete quantum system used in our algorithm comprises $n+1$ qubits, with $n$ qubits for executing the oracle (playing the role of the system of interest in the modular value measurement scheme) and one for the readout process (playing the role of the ancillary meter component). The algorithm consists of three main parts: (1) pre-selection of the oracle input state and preparation of the ancilla's state, (2) controlled oracle application, and (3) post-selection on the system to the final state and readout of the oracle to extract information related to the modular value. Finally, the measurement of the modular value unfolds by measuring the meter qubit (in case of successful post-selection).
\\
To begin, we take the oracle subsystem's $n$ qubits to the pre-selected state
\begin{equation}
\ket{\psi_i}=\ket{\psi_i^{1}}\otimes\ket{\psi_i^{2}}\otimes...\otimes\ket {\psi_i^{n}}, 
\end{equation}
by using unitary operators. The initial state of the meter, denoted as the projector $\hat{\Pi}_m$, is defined as
\begin{equation}
\label{eq:rho_m}
\hat{\Pi}_{\vec{m}}=\frac{1}{2}\left(\hat{I}+\Vec{m}\cdot\hat{\Vec{\sigma}}\right),
\end{equation}
where $\Vec{m}$ represents the state on the Bloch sphere, and $\hat{I}$ is the identity operator. To obtain this initial state, we employ a unitary operator.\\
Subsequently, we apply the following nonlocal unitary operator to the entire system, comprising the oracle subsystem and the meter:
\begin{equation}
\hat{U}_{\text{GATE}}=\hat{\Pi}_{\vec{r}}\otimes\hat{I}+\hat{\Pi}_{-\vec{r}}\otimes\hat{U}_O.
\end{equation}
In this expression, $\hat{U}_O$ represents the oracle employed in the experiment, while $\hat{\Pi}_{\vec{r}}$ and $\hat{\Pi}_{-\vec{r}}$ are projectors applied to the ancillary space. These projectors can be expressed in terms of the vector $\pm\Vec{r}$ on the Bloch sphere as follows:
\begin{equation}
\label{eq:pi_r}
\hat{\Pi}_{\pm \vec{r}}=\frac{1}{2}\left(\hat{I}\pm\Vec{r}\cdot\hat{\Vec{\sigma}} \right).
\end{equation}
The unitary operator $\hat{U}_{\text{GATE}}$ facilitates a controlled evolution of the oracle subsystem. When the ancillary projector is $\hat{\Pi}_{\vec{r}}$, the oracle subsystem does not evolve as it undergoes only the identity operation. Conversely, when the ancillary projector is $\hat{\Pi}_{-\vec{r}}$, the oracle subsystem experiences the application of the oracle operator $\hat{U}_O$. In the case of an ancillary state corresponding to a linear combination of these two scenarios, the superposed effect of both operators is applied to the oracle subsystem.\\
To complete the procedure, we post-select the oracle subsystem qubits to $\ket{\psi_f}$. Finally, by measuring the spin operator in the direction $\Vec{q}$,
\begin{equation}
\label{eq:sigma_q}
\hat{\sigma}_q=\hat{\Pi}_{+\vec{q}}-\hat{\Pi}_{-\vec{q}},
\end{equation} 
we extract the information on the modular value upon successful post-selection.
After some mathematical development shown in \cite{cormann2016revealing}, the joint probabilities of finding the $n$ qubits of the system in the composite state in $\ket{\psi_f}$ and the meter in $\pm q$ are computed from
\begin{equation}
P^{\pm \vec{q}}_{joint}=\text{Tr}\left[\left(\ket{\psi_f}\bra{\psi_f}\otimes\hat{\Pi}_{\pm \vec{q}}\right)\hat{\rho}\right],
\end{equation}
where $\hat{\rho}$ is the state of the full system after the controlled application of the oracle. When applying the quantum eraser condition $\Vec{r}\cdot\Vec{q}=0$ on the meter measurement basis\footnote{Selecting two quantum states that are orthogonal on the Bloch sphere, expressed as $\Vec{r}\cdot\Vec{q}=0$, maximizes the interference between different pathways during the meter measurement. In this configuration, the nonlocal gate action arises as a superposition of both $\hat{U}_O$ and $\hat{I}$, by causing a loss of information about the gate action. This particular circumstance is referred to as the ``quantum eraser'' condition, a term frequently employed in interferometer experiments to eliminate the knowledge of a particle's path. For instance, by applying a linear polarizer can erase the information regarding whether the polarization is circular left or right. All linear polarizations lie orthogonal on the Poincaré sphere to both left and right circular polarizations.}, the joint probabilities take the form
\begin{eqnarray}
   \label{eq:joint_probabilities}
   &&P^{\pm \vec{q}}_{joint}=\frac{1}{4}\{\left(1+\Vec{r}\cdot\Vec{m}\right)\lvert\bra{\psi_f}\ket{\psi_i}\rvert^2 \\ \nonumber
   &+&\left(1-\Vec{r}\cdot\Vec{m}\right)\lvert\bra{\psi_f}\hat{U}_O\ket{\psi_i}\rvert^2\\\nonumber
   &\pm&2\left(\Vec{m}\cdot\Vec{q}\right)\text{Re}\left(\bra{\psi_f}\hat{U}_O\ket{\psi_i}\bra{\psi_i}\ket{\psi_f}\right) \\ \nonumber
   &\pm& 2\left[\left(\Vec{r}\times\Vec{m}\right)\cdot\Vec{q}\right]\text{Im}\left(\bra{\psi_f}\hat{U}_O\ket{\psi_i}\bra{\psi_i}\ket{\psi_f}\right)\}.\nonumber
\end{eqnarray}
The expected value of the spin operator in the meter is \cite{cormann2016revealing},
\begin{eqnarray}\label{eq:expected_value_spin_operator}
\bar{\sigma}_q^m&=&2 \frac{\left(\Vec{m}\cdot\Vec{q}\right)\text{Re} O_m+\left[\left(\Vec{r}\times\Vec{m} \right)\cdot\Vec{q} \right]\text{Im} O_m}{\left(1+\Vec{r}\cdot\Vec{m} \right)+\left(1- \Vec{r}\cdot\Vec{m}\right)|O_m|^2} \\ \nonumber
&=&\frac{P_{\text{joint}}^{+\vec{q}}-P_{\text{joint}}^{-\vec{q}}}{P_{\text{joint}}^{+\vec{q}}+P_{\text{joint}}^{-\vec{q}}}, 
\end{eqnarray}
where the modular value is
\begin{equation}
O_m=\frac{\bra{\psi_f}\hat{U}_O\ket{\psi_i}}{\bra{\psi_f}\ket{\psi_i}}.
\end{equation} 
The qubit meter measurement extracts thus information on the modular value of the oracle, which in turn provides 
information relevant to solving the problem at hand, as will be illustrated in the following two sections. We note in particular that by choosing appropriate relative orientations of the meter directions $\vec{m}$, $\vec{r}$, and $\vec{q}$, it is possible to select the real of the imaginary part of the modular value in the numerator of expression (\ref{eq:expected_value_spin_operator}).\\
In full generality, we can already state that to determine the required number of repetitions of the algorithm to obtain enough statistics to solve the initial problem, we need to evaluate the probability of post-selection, which is denoted at first order as
\begin{equation}
p=|\bra{\psi_f}\ket{\psi_i}|^2,
\end{equation}
and the visibility that measures the contrast between joint probabilities that we measure, which is given by
\begin{equation}
\label{eq:visibility_algorithm}
V=\frac{P_{max}-P_{min}}{P_{max}+P_{min}}.
\end{equation}
$P_{\text{min}}$ represents the minimum joint probability between $P_{\text{joint}}^{+\vec{q}}$ and $P_{\text{joint}}^{-\vec{q}}$, while $P_{\text{max}}$ corresponds to the maximum joint probability. In each particular case, the determination of the maximum and minimum joint probabilities is essential and can be achieved by considering the expression provided in Eq.~\ref{eq:joint_probabilities}. When employing this approach, there exist various degrees of freedom that can be optimized based on specific experimental conditions. These include selecting the pre- and post-selected states, choosing the meter state, and determining the operator to be measured in the meter. Additionally, alternative experimental approaches can be explored, which may lead to the emergence of modular values and the development of novel methods for executing quantum algorithms.
\section{Standard quantum Deutsch-Jozsa algorithm}
\begin{figure*} [th!]
\centering 
\includegraphics[width=0.7\textwidth]{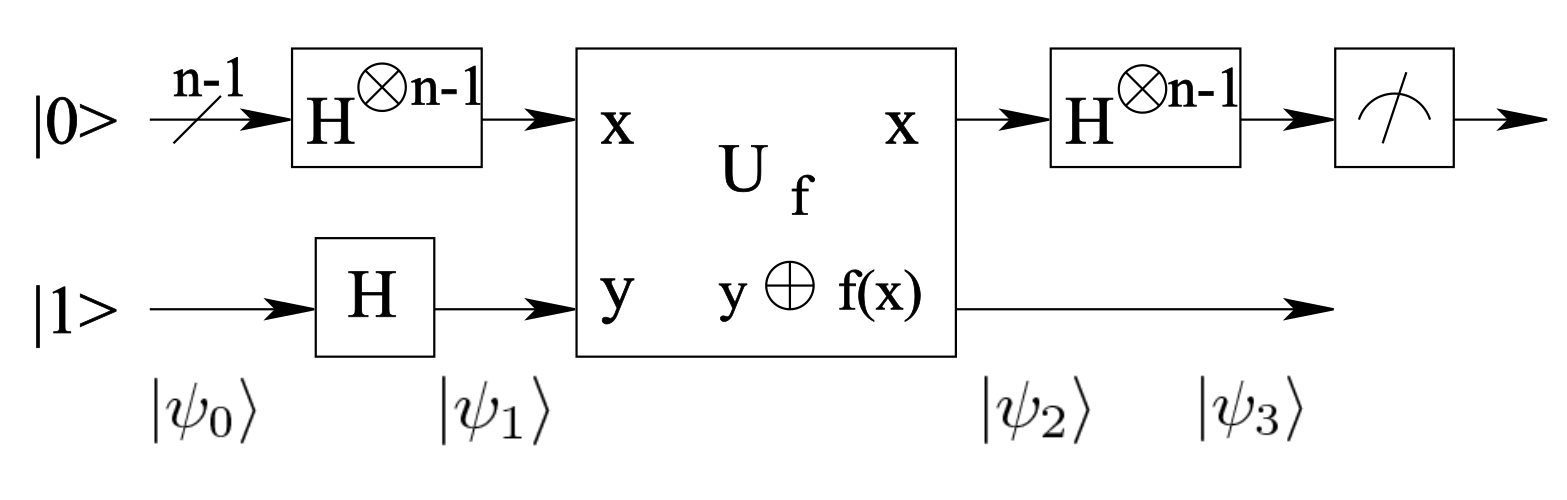}
  \caption{Deutsch-Jozsa quantum circuit, where $U_f$ is the oracle, $H$ is the Hadamard gate, and the states $\ket{\psi_0}$, $\ket{\psi_1}$, $\ket{\psi_2}$, and $\ket{\psi_3}$ are detailed in the text. Modified from Peplm, https://commons.wikimedia.org/wiki/File:Deutsch-
  Jozsa-algorithm-quantum-circuit.png, accessed on the 9th of May 2023. \label{fig:deutsch_jozsa_qauntum_circuit}}
\end{figure*}

The Deutsch-Jozsa algorithm aims to determine whether a given function $f\left(x\right):\{0,1\}^{n-1}\rightarrow\{0,1\}$ is constant or balanced. A constant function assigns the same value (either $0$ or 1) to all the inputs, whereas a balanced function assigns the value $1$ to exactly half of the inputs and the value $0$ to the other half. Table~\ref{tab:possible_combination_balanced_and_constant} shows the different possibilities for the Deutsch-Jozsa algorithm for a function $f\left(x\right)$ with two-bit inputs (there are thus four possible inputs with two bits: 00, 01, 10, and 11). 
\begin{table}[th!]
\centering
\caption*{Balanced} 
 \begin{tabular}{|| c| c| c| c ||} 
 \hline
 $f(00)$ & $f(01)$ & $f(10)$ & $f(11)$ \\ [0.5ex] 
 \hline\hline
0 & 0 & 1 & 1  \\ 
 \hline
0 & 1 & 0 & 1  \\
 \hline
0 & 1 & 1 & 0   \\
 \hline
1 & 1 & 0 & 0   \\
 \hline
1 & 0 & 1 & 0   \\
\hline
1 & 0 & 0 & 1   \\[1ex] 
 \hline
\end{tabular}
\caption*{Constant} 
 \begin{tabular}{|| c| c| c| c ||} 
 \hline
 $f(00)$ & $f(01)$ & $f(10)$ & $f(11)$ \\ [0.5ex] 
 \hline\hline
0 & 0 & 0 & 0  \\ 
 \hline
1 & 1 & 1 & 1  \\ [1ex] 
 \hline
\end{tabular}
\caption{\label{tab:possible_combination_balanced_and_constant} 
All the possible combinations of two-bit inputs and single-bit outputs that result in either a constant or a balanced function.}
\end{table}\\
In 1992, Deutsch and Jozsa introduced a groundbreaking quantum circuit that solves this problem in a single step, providing a significant advantage over the classical counterpart. In the worst-case scenario, the classical approach requires $2^{n-1}$ repetitions, where $2^{n-1}$ represents the total number of different inputs for the function. However, by leveraging qubits, we can exploit the fact that $2^{n-1}$ corresponds to the dimensionality of the space formed by $n-1$ qubits. Thus, we can effectively employ $n-1$ qubits to generate the entire set of $2^{n-1}$ inputs. In stark contrast to classical approaches, the quantum circuit designed by Deutsch and Jozsa achieves the same result efficiently. The quantum circuit is depicted in Fig.~\ref{fig:deutsch_jozsa_qauntum_circuit}, showcasing its elegance and simplicity. To tackle the problem with $2^{n-1}$ inputs, the algorithm mandates the utilization of $n$ qubits.
To address a problem involving a function with all possible combinations of $n-1$ qubits as input, it is necessary to utilize a total of $n$ qubits. The first $n-1$ qubits are initialized in the state $\ket{0}$, while the remaining qubit is initialized in the state $\ket{1}$. This initialization results in the following state:
\begin{equation}
\ket{\psi_0}=\ket{0}^{\otimes (n-1)}\ket{1} 
\end{equation}
where $\ket{0}^{\otimes (n-1)}$ is a compact notation for the tensor product of $n-1$ state kets $\ket{0}$, i. e. $\ket{0}\otimes\ket{0}\hdots\otimes\ket{0}$, and the basis has been chosen as 
\begin{equation}
\ket{0}=\begin{pmatrix}
1\\
0
\end{pmatrix}
\hspace{1 cm}
\ket{1}=\begin{pmatrix}
0\\
1
\end{pmatrix}.
\end{equation}
The first step consists of applying the Hadamard gate, $\hat{H}$, 
\begin{equation}
\label{eq:hadamard_gate}
\hat{H}=\frac{1}{\sqrt{2}}\begin{pmatrix}
1 & 1 \\
1 & -1 \\
\end{pmatrix}
\end{equation}
to all $n-1$ qubits. The state after the transformation is
\begin{eqnarray}
 \ket{\psi_1}&=&\frac{1}{\sqrt{2^{n}}}\left(\ket{0}+\ket{1}\right)^{\otimes n-1}\left(\ket{0}-\ket{1}\right)\\ \nonumber
 &=&\sum_{x\in\{0,1\}^{n-1}}\frac{1}{\sqrt{2^{n}}}\ket{x}\left(\ket{0}-\ket{1}\right),
\end{eqnarray}
where $\ket{x}$ are all the $2^{n-1}$ states of the basis with $n-1$ qubits.\\ 
After applying the Hadamard gates, the oracle is applied to the $n$ qubits. The oracle acts as
\begin{equation}
\hat{U}_f\ket{x,y}\rightarrow \ket{x,y\oplus f\left(x\right)},
\end{equation}
where $\oplus$ represents the addition modulo 2, that is $0$ if both terms are equal and $1$ if both terms are different 
\begin{eqnarray}
0\oplus 0&=&0 \\ \nonumber
0 \oplus 1&=&1 \\ \nonumber
1 \oplus 0&=&1 \\ \nonumber
1 \oplus 1&=&0.
\end{eqnarray}
The resulting state after the application of the oracle is:
\begin{eqnarray}
\label{eq:psi_2_deutsch_jozsa}
\nonumber\ket{\psi_2}&=&\sum_{x\in\{0,1\}^{n-1}}\frac{1}{\sqrt{2^{n}}}\ket{x}\left(\ket{0\oplus f\left(x\right)}-\ket{1\oplus f\left(x\right)}\right) \\
&=&\sum_{x\in\{0,1\}^{n-1}}\frac{\left(-1\right)^{f\left(x\right)}}{\sqrt{2^{n}}}\ket{x}\left(\ket{0}-\ket{1}\right).
\end{eqnarray}
In Eq.~\ref{eq:psi_2_deutsch_jozsa}, we make use of the property that $\ket{0\oplus f\left(x\right)}=\ket{f\left(x\right)}$, while $\ket{1\oplus f\left(x\right)}=\ket{1-f\left(x\right)}$.\\
Subsequently, Hadamard gates, denoted as $\hat{H}^{\otimes \left(n-1\right)}$, are applied to the first $n-1$ qubits. The application of $n-1$ Hadamard gates to an $n-1$-qubit state can be expressed as:
\begin{eqnarray} 
&&\hat{H}^{\otimes n-1}\ket{x_1,\hdots,x_{n-1}} \\ \nonumber
&=&\frac{\sum_{z_1,\hdots,z_{n-1}}\left(-1\right)^{x_1z_1+\hdots+x_{n-1}z_{n-1}}\ket{z_1, \hdots, z_{n-1}}}{\sqrt{2^{n-1}}},
\end{eqnarray}
where $x_i$ are the components of the state $\ket{x}$, while $z_i$ are the ones of the state $\ket{z}$, hence every component is either a 0 or a 1.\\
The notation can be simplified as
\begin{equation}
\hat{H}^{\otimes \left(n-1\right)}\ket{x}=\frac{\sum_z\left(-1\right)^{x\cdot z}\ket{z}}{\sqrt{2^{n-1}}},
\end{equation}
where $x\cdot z$ is the inner product of x and z, modulo 2. Finally, the state $\ket{\psi_3}$ is obtained after applying the Hadamard gate to the first $n-1$ qubits,
\begin{equation}
\ket{\psi_3}=\sum_{z_1,\hdots,z_{n-1}}\sum_{x_1,\hdots,z_{n-1}}\frac{1}{\sqrt{2^{n}}}\left(-1\right)^{x\cdot z+f\left(x\right)}\ket{z}\left(\ket{0}-\ket{1}\right)
\end{equation}
Eventually, if we measure in the $\{\ket{0}, \ket{1}\}^{n-1}$ basis the $n-1$ first qubits, the probability of obtaining the state $\ket{0}^{\otimes \left(n-1\right)}$ would be, 
\begin{equation}
\Bigg|\frac{1}{2^{n-1}}\sum_{x=0}^{2^{n-1}-1}\left(-1\right)^{f\left(x\right)}\Bigg|.
\end{equation}
Consequently, if the measurement result is $\ket{0}^{\otimes (n-1)}$, the function is constant, whereas if the measurement result is different from $\ket{0}^{\otimes (n-1)}$, the function is balanced. By applying the algorithm once, we can determine with certainty whether the function is balanced or constant.\\
Although the Deutsch-Jozsa algorithm does not have any real-world applications, it was one of the first quantum algorithms to demonstrate a clear advantage over its classical counterpart.

\section{Deutsch-Jozsa-inspired algorithm by using modular values}
In the previous section, we discussed the conventional method for solving the Deutsch-Jozsa problem. In this section, we explore an alternative approach by calculating the modular value of the standard oracle operator $\hat{U}_O$. When solving the problem with $2^{n-1}$ possible inputs, the oracle is applied to $n$ qubits. As a result, the pre- and post-selected states in our procedure correspond to $n$-qubit states. To extract the modular value, an additional qubit, known as the ancilla qubit, is required. Therefore, a total of $n+1$ qubits are needed in our approach. The application of the oracle can be expressed as
\begin{equation}
    \hat{U}_O\left(\ket{x}\ket{y}\right)=\ket{x}\ket{y\oplus f\left(x\right)}.
    \label{eq:unitary_operator}
\end{equation}
The algorithm can be implemented by calculating the modular value of the oracle in systems of any dimension.\\
The Deutsch-Jozsa problem for a function of $4$ different inputs (all possible values of two bits) requires two qubits for the oracle, and thus three qubits in total in order to evaluate the modular value of the oracle, as seen in the previous section. In the modular value approach, the oracle subsystem spans thus two qubits, while we need an extra qubit to act as the ancilla. With these dimensions, the chosen pre- and post-selected states are:
\begin{eqnarray}
\label{eq:pre_and_post_three_qubits}
\ket{\psi_i}&=&\frac{1}{\sqrt{8}}\left(\ket{0}+\ket{1}\right)\otimes\left(\ket{0}+\ket{1}\right) \\ \nonumber
&\otimes&\left(\ket{0}-\ket{1}\right)\\ \nonumber
&=&\frac{1}{\sqrt{8}}(\ket{000}+\ket{010}+\ket{100}+\ket{110}\\ \nonumber
&-&\ket{001}-\ket{011}-\ket{101}-\ket{111})\\\nonumber
\ket{\psi_f}&=&\frac{1}{\sqrt{11}}(-2i\ket{000}-i\ket{010}+i\ket{001} \\ \nonumber
&+&i\ket{011}+\ket{100}+\ket{110}-\ket{101}-\ket{111}).\nonumber
\end{eqnarray}
The probability of post-selection in this case is given by $p=|\bra{\psi_f}\ket{\psi_i}|^2=\frac{41}{88}\approx 0.47$, which is slightly below one-half. Consequently, on average, two attempts are needed to achieve successful post-selection. When the function $f\left(x\right):\{0,1\}^{n-1}\rightarrow\{0,1\}$ is constant, meaning that $f\left(x\right)$ is either $0$ or $1$ for all $x$, the numerator of the modular value is equal to the denominator or its opposite. Thus, the modular value is either $1$ or $-1$, resulting in a purely real value. On the other hand, when the function $f\left(x\right):\{0,1\}^n\rightarrow\{0,1\}$ is balanced, with half of the terms $f\left(x\right)$ equal to $1$ and the other half equal to $0$, the modular values possess an imaginary part for all possible balanced combinations of $f\left(x\right)$.
\begin{table*}[ht]
\centering
 \begin{tabular}{|| c| c| c| c| c|c|c||}
 \hline
 $f(00)$ & $f(01)$ & $f(10)$ & $f(11)$ & $O_m$ & $\text{Re}\left(O_m\right)$ & $\text{Im}\left(O_m\right)$\\ [0.5ex] 
 \hline\hline
0 & 0 & 1 & 1 & $\frac{5i-4}{5i+4}$ & $\frac{9}{41}$ & $\frac{40}{41}$\\ 
 \hline
0 & 1 & 0 & 1 & $\frac{i}{5i+4}$ & $\frac{5}{41}$ & $\frac{4}{41}$\\
 \hline
0 & 1 & 1 & 0 & $\frac{i}{5i+4}$ & $\frac{5}{41}$ & $\frac{4}{41}$\\
 \hline
1 & 1 & 0 & 0  & $\frac{-5i+4}{5i+4}$ & $-\frac{9}{41}$ & $-\frac{40}{41}$\\ 
 \hline
1 & 0 & 1 & 0 & $\frac{-i}{5i+4}$ & $-\frac{5}{41}$ & $-\frac{4}{41}$\\
\hline
1 & 0 & 0 & 1 & $\frac{-i}{5i+4}$ & $-\frac{5}{41}$ & $-\frac{4}{41}$\\ [1ex]
\hline
\end{tabular}
\caption{\label{tab:combination_balanced_three_qubits} All the possible combinations to have a balanced function with three qubits and the corresponding modular values.}
\end{table*}\\
In Table~\ref{tab:combination_balanced_three_qubits}, all balanced possibilities using three qubits (two qubit inputs and one ancilla) are presented.
All modular values have an imaginary part, which allows for discrimination between balanced and constant functions by measuring this part. This process is easily generalized to the algorithm with $n+1$ qubits (the $n$ qubits of the Deutsch-Jozsa algorithm plus the ancilla). In this case, the pre-selected state is the natural generalization of the previous case
\begin{equation}
    \ket{\psi_i}=\frac{1}{\sqrt{2^n}}\left(\ket{0}+\ket{1}\right)^{\otimes \left(n-1\right)}\left(\ket{0}-\ket{1}\right).
     \label{eq:pre-selection_general_deutsch_jozsa}
\end{equation}
The post-selected state is generalized from the three-qubit case,
\begin{eqnarray}
    \label{eq:post-selection_general_deutsch_jozsa}
    \ket{\psi_f}&=&\frac{1}{\sqrt{2^n+3}}(-2i\ket{00...0}+i\ket{00...0...01} \\ \nonumber
    &+&(-i\sum_{j=2}^{2^{n-2}} \underbrace{\ket{0a_2...a_{(2^{n-1})}}_j}_{\begin{subarray}{l}\text{$(2^{n-1})/2-1$ first terms with $n-1$}\\\text{qubits without including $\ket{000...0}$}\end{subarray}}\\ \nonumber 
    &+&\sum_{k=2^{n-2}+1}^{2^{n-1}}\underbrace{\ket{1b_2...b_{(2^{n-1})}}_k}_\text{$(2^{n-1})/2$ last terms with $n-1$ qubits})\\ \nonumber
    &\times&\left(\ket{0}-\ket{1}\right)) 
\end{eqnarray}
The probability of post-selection using the states Eq.~\ref{eq:pre-selection_general_deutsch_jozsa} and Eq.~\ref{eq:post-selection_general_deutsch_jozsa} is
\begin{equation}
p=\frac{2^n\left( 1+2^{n-1}\right)+1}{2^n\left(3+2^n\right)}=\frac{2^{2n-1}+2^n+1}{2^{2n}+3\times 2^n},
\end{equation}
whose limit when $n$ tends to infinity is $\frac{1}{2}$. Therefore, on average, two attempts are required for successful post-selection. In the case where the function $f\left(x\right):\{0,1\}^{n-1}\rightarrow\{0,1\}$ is constant, the numerator and denominator of the modular value take the same value or the opposite value, resulting in a modular value of either $1$ or $-1$. Conversely, when the function is balanced, the modular value presents always an imaginary part. Discrimination between a real and complex modular value can be accomplished by selecting appropriate measurement directions (i.e., vector $\vec{q}$) for the ancillary component. Indeed, the measurement results provided by Eq.~\ref{eq:expected_value_spin_operator} depend on both the real and imaginary parts of the modular value. However, by choosing vectors $\Vec{m}$ and $\Vec{q}$ that are perpendicular, the real part does not contribute to the numerator (because $\Vec{m}\cdot\Vec{q}=0$ in Eq.~\ref{eq:expected_value_spin_operator}), so that a non-zero measurement result indicates that the modular value possesses a non-zero imaginary part. Conversely, by choosing vectors such that $\left(\Vec{r}\times\Vec{m}\right)\cdot\Vec{q}=0$, only the real part contributes to the numerator providing the measurement results. Furthermore, through the meticulous selection of appropriate pre- and post-selected states, one can ascertain the specific way in which the function is balanced among the various possibilities (each row of the balanced table in Table~\ref{tab:possible_combination_balanced_and_constant}). In such cases, the modular value associated with the oracle would vary for each distinct manner of achieving a balanced function, corresponding to each row in the balanced table presented in Table~\ref{tab:possible_combination_balanced_and_constant}.
\section{Choice of pre- and post-selected states in the Deutsch-Jozsa-inspired algorithm by using modular values}
\begin{figure} [t!]
\centering 
\includegraphics[width=0.5\textwidth]{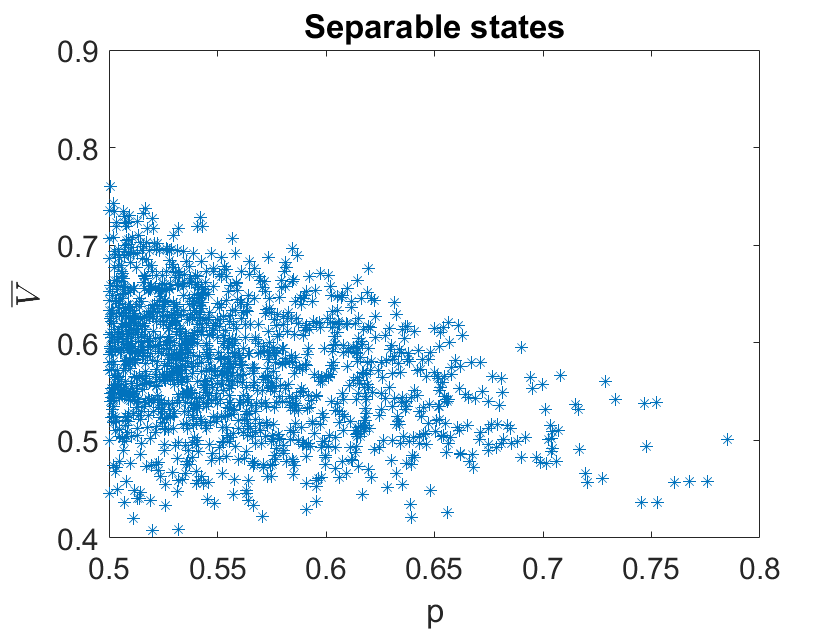}
  \caption{Average visibility as a function of the probability of post-selection in the Deutsch-Jozsa problem for the 2 qubits case. All pre- and post-selected states are separable two-qubit states and are generated randomly. Our criteria for selecting these states were to ensure that, for each balanced function, the imaginary part of the weak value of the oracle is greater than $0.1$, each individual visibility is greater than $0.4$, and the probability of post-selection is greater than $0.5$. By imposing these conditions, we aim to strike a balance between a high post-selection probability, and strong visibility to ensure the ability to distinguish between constant and balanced functions.\label{fig:visibility_probability_separable}}
\end{figure}

\begin{figure} [t!]
\centering 
\includegraphics[width=0.5\textwidth]{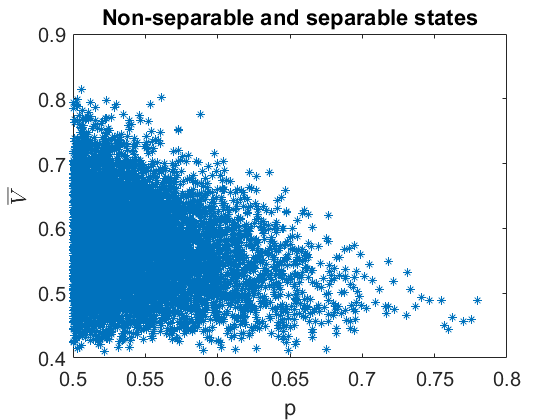}
  \caption{Average visibility as a function of the probability of post-selection in the Deutsch-Jozsa problem for the 2 qubits case.
All pre- and post-selected states are generated randomly without constraint (a typical state will thus be non-separable). Average visibility in terms of the probability of post-selection for randomly generated two-qubit non-separable and separable pre- and post-selected states in the context of the Deutsch-Jozsa problem. For this analysis, we have imposed a minimum threshold of $0.1$ for the imaginary part of each weak value of the oracle, considering all balanced functions. Additionally, each individual visibility is required to exceed $0.4$, and the probability of post-selection must be greater than $0.5$.\label{fig:visibility_probability_non_separable}}
\end{figure}

In the previous section, we introduced the pre- and post-selected states for solving the Deutsch-Jozsa problem using modular values, as described in Eq.~\ref{eq:pre-selection_general_deutsch_jozsa} and Eq.~\ref{eq:post-selection_general_deutsch_jozsa}. However, one might question the rationale behind this particular choice. Our objective is to find two quantum states (pre-selected and post-selected) that yield a non-zero imaginary part of the weak value of the oracle for all possible balanced functions $f(x)$, as given in Table~\ref{tab:possible_combination_balanced_and_constant}. Ideally, we aim to maximize this imaginary part while minimizing the real part of the modular value, in order to enhance the visibility defined in Eq.~\ref{eq:visibility_algorithm}. Additionally, we strive to maximize the probability of successful post-selection. Hence, we are confronted with an optimization problem.\\
In order to explore the entire quantum space, we conducted a random mapping of the complete state space for the pre-selected and post-selected states, both consisting of four-level states (two qubits). The pseudorandom numbers used were integer values drawn from a discrete uniform distribution. Fig.~\ref{fig:visibility_probability_separable} showcases the resulting average visibility of all combinations of the balanced function, which are further detailed in Table~\ref{tab:possible_combination_balanced_and_constant}, as a function of the probability of post-selection. In this particular case, we have chosen to focus on separable states, as they are easier to generate using quantum gates, such as those available in the IBM quantum computer. As depicted in the plot, there is a trade-off that needs to be considered. On one hand, we can achieve a very high probability of post-selection, but at the cost of lower visibility. On the other hand, we can enhance the visibility, but with a decrease in the probability of successful post-selection.
\\
Alternatively, we can explore the space of non-separable states (including separable ones) as a potential solution. In Fig.~\ref{fig:visibility_probability_non_separable}, we present a denser plot where we consider both separable and non-separable states. This broader range of states offers more possibilities for achieving intermediate values of average visibility in relation to the probability of post-selection. It was expected that including non-separable states would provide additional options beyond the separable ones. When aiming for low visibility (around $0.4$) and high post-selection probability (around $0.8$), the benefit of incorporating non-separable states is limited. However, when targeting a lower probability of post-selection (around $0.5$), we observe points on the plot that yield improved average visibility values exceeding $0.8$.
Interestingly, we observe the emergence of a distinct trend in the form of a dividing line, which seems to move up when non-separable states are considered. This suggests that there is a clear boundary beyond which points are not found above the line. It is worth noting that by choosing non-separable states introduces additional complexity in generating such states on a quantum computer. This typically involves utilizing multi-qubit gates such as the CNOT or the TOFFOLI gates. Despite the challenges involved, non-separable states offer potential advantages, as demonstrated by the improved average visibility and probability of post-selection in certain regions of the plot. Given the trade-off between selecting separable and non-separable states, we opted for simplicity in the implementation of the Deutsch-Jozsa problem on the IBM quantum computer in the subsequent section. We specifically chose straightforward pre- and post-selected states that enable differentiation between only one particular balanced function, and constant functions. This decision was motivated by the difficulty of implementing more complex states using the available gates.

\section{Implementation of the the Deutsch-Jozsa-inspired algorithm by using modular values in the IBM quantum computer}
\begin{figure*} [th!]
\centering 
\includegraphics[width=0.7\textwidth]{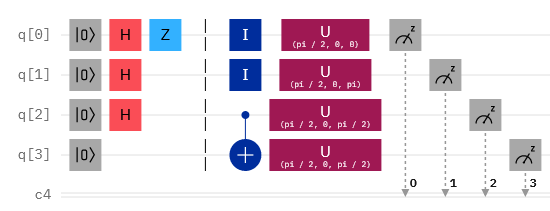}
  \caption{The circuit diagram illustrates the implementation of the modular value algorithm for the  Deutsch-Jozsa problem for the balanced case, using a $3$-qubit system and one ancilla qubit. The gates applied to each qubit are $\hat{U}\left(\pi/2,0,\pi\right)$ to qubit $0$, $\hat{U}\left(\pi/2,0,\pi\right)$ to qubit $1$, $\hat{U}\left(\pi/2,0,\pi/2\right)$ to qubit $2$, and $\hat{U}\left(\pi/2,0,\pi/2\right)$ to qubit $3$. Figure produced with https://quantum-computing.ibm.com.\label{fig:scheme_circuit_balanced}}
\end{figure*}

\begin{figure} [th!]
\centering 
\includegraphics[width=0.5\textwidth]{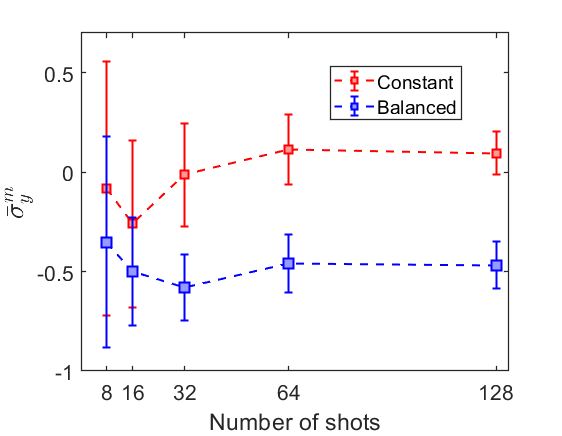}
  \caption{Mean $\hat{\sigma}_y$ operator measurement outcomes for both the constant and balanced cases across a range of shots ($8$, $16$, $32$, $64$, and $128$): experiment run in the ibmq\_quito quantum computer. The entire procedure has been iterated a total of $12$ times for each distinct run count. Furthermore, to provide a comprehensive representation, the standard deviation has been incorporated in the form of error bars. \label{fig:results_ibm_computer}}
\end{figure}

\begin{figure} [th!]
\centering 
\includegraphics[width=0.5\textwidth]{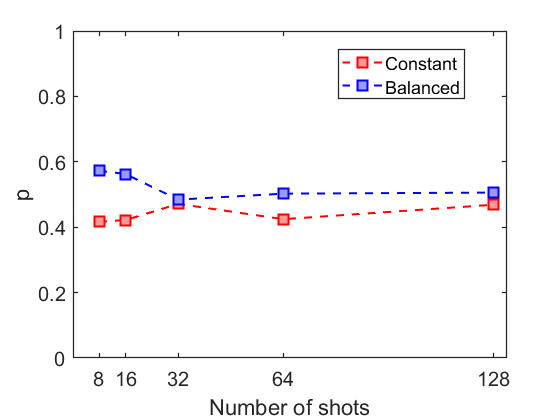}
  \caption{Experimental average post-selection rate, p, for varying number of shots: $8$, $16$, $32$, $64$, and $128$, encompassing both the constant and balanced scenarios. Experiment run in the ibmq\_quito quantum computer. \label{fig:post-selected_rate}}
\end{figure}

\begin{figure} [th!]
\centering 
\includegraphics[width=0.5\textwidth]{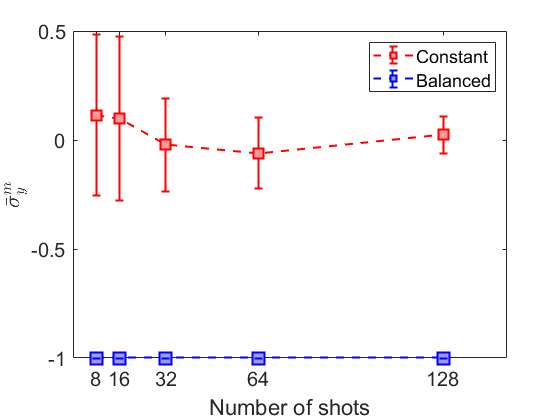}
  \caption{Mean $\hat{\sigma}_y$ operator measurement outcomes for both the constant and balanced cases across a range of shots ($8$, $16$, $32$, $64$, and $128$): experiment run in the IBM simulator\_mps. The entire procedure has been iterated a total of $12$ times for each distinct run count. Furthermore, to provide a comprehensive representation, the standard deviation has been incorporated in the form of error bars. \label{fig:results_ibm_computer_simulator}}
\end{figure}

\begin{figure} [th!]
\centering 
\includegraphics[width=0.5\textwidth]{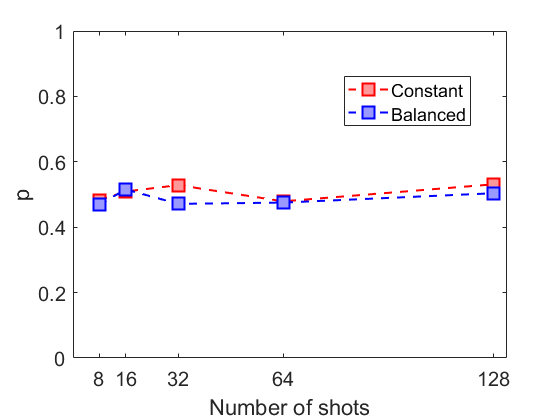}
  \caption{Experimental average post-selection rate, p, for varying number of shots: $8$, $16$, $32$, $64$, and $128$, encompassing both the constant and balanced scenarios. Experiment run in the IBM simulator\_mps. \label{fig:post-selected_rate_simulator}}
\end{figure}

To experimentally implement our algorithm tackling the Deutsch-Jozsa problem, we can follow the scheme shown in Fig.~\ref{fig:scheme_model}. In this approach, we need to select the initial state of the meter, denoted by $\ket{m}$ (Eq.~\ref{eq:rho_m}), the projectors $\hat{\Pi}_{\pm\Vec{r}}$ (Eq.~\ref{eq:pi_r}), and the final state measured in the meter, denoted by $\ket{q}$ (Eq.~\ref{eq:sigma_q}). We should impose $\Vec{r}\cdot\Vec{q}=0$ to satisfy the quantum eraser condition. Additionally, we choose to enforce $\Vec{m}\cdot\Vec{q}=0$ in order to have direct access to the imaginary part of the modular value (see Eq.~\ref{eq:expected_value_spin_operator}). The initial meter state corresponds thus to
\begin{equation}
\hat{\Pi}_{\vec{m}}=\frac{1}{2}\left(\hat{I}+\hat{\sigma}_z\right),
\end{equation}
where $\Vec{m}=\left(0,0,1\right)$. The projector controlling the application of the oracle is 
\begin{equation}
\hat{\Pi}_{\vec{r}}=\frac{1}{2}\left(\hat{I}+\hat{\sigma}_x \right),
\end{equation}
where $\Vec{r}=\left(1,0,0\right)$. The observable to measure in the meter is $\hat{\sigma}_y$, with $\Vec{q}=\left(0,1,0\right)$.\\
Applying the selected states, the average measured value of $\hat{\sigma}_y$ in the probe is 
\begin{equation}
    \bar{\sigma}_y^m=\frac{-2\ \text{Im}\left(O_m\right)}{1+|O_m|^2}.
\end{equation}
Taking into account the joint probability values in this scenario, the visibility is given by
\begin{equation}
V=\frac{2\ |\text{Im}\left(O_m\right)|}{1+|\text{Im}\left(O_m\right)|^2+|\text{Re}\left(O_m\right)|^2}. 
\end{equation}
It is desirable to achieve a high visibility to distinguish in fewer repetitions of the experiment the real part from the imaginary one. In particular, the maximum visibility $V=1$ is reached for $\text{Im}\left(O_m\right)=\pm 1$ and $\text{Re}\left(O_m\right)=0$. The closer the real part to 0, the better; the closer the absolute value of the imaginary part to $1$, the better. This is equivalent to maximizing $\frac{2x}{x^2+y^2+1}$ with respect to $x$ and $y$. For the case of the Deutsch-Jozsa problem with two qubits, $f\left(x\right):\{0,1\}^2\rightarrow \{0,1\}$ shown in Table~\ref{tab:combination_balanced_three_qubits}, the first and third possibilities to balance the system have a good visibility, while the others do not. These values can be optimized at the expense of losing some visibility on the first and third cases. The number of times the algorithm should be executed, named shots in the jargon, depends on the level of uncertainty one is willing to accept.\\
We ran the modular value algorithm to solve the Deutsch-Jozsa problem on the Quito IBM quantum computer (ibmq\_quito), by using modular values, for one constant and one balanced case. The pre-selected state is chosen as described in Eq.~\ref{eq:pre_and_post_three_qubits}, and we simplified the post-selected state by choosing it as follows
\begin{eqnarray}
\label{eq:post_selected_state_IBM}
\ket{\psi_f}&=&\frac{1}{\sqrt{8}}(\ket{000}+\ket{010}+i\ket{100} \\ \nonumber
&+&i\ket{110}-\ket{001}-\ket{011}-i\ket{101}-i\ket{111}).
\end{eqnarray}
The latter state has been chosen, because it can be easily obtained through simple gates on the IBM quantum computer. Let us however observe that in this case, only one type of balanced function can be discriminated from the constant case, namely a single raw in the top Table~\ref{tab:possible_combination_balanced_and_constant}. In a next step, it would be desirable to implement the post-selected state expressed in Eq.~\ref{eq:pre_and_post_three_qubits}. The probability of post-selection by using the state described in Eq.~\ref{eq:post_selected_state_IBM} is $p=\frac{1}{2}$. The states $\ket{m}$, $\ket{r}$, and $\ket{q}$ were chosen as described earlier. 
\begin{figure*}
\centering 
\includegraphics[width=0.7\textwidth]{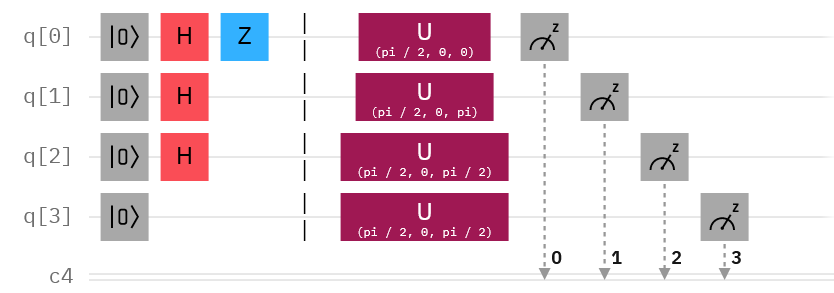}
  \caption{Circuit diagram for the modular value algorithm for the  Deutsch-Jozsa problem with a constant function using $3$ qubits and one ancilla qubit. The circuit applies the unitary operator $\hat{U}\left(\frac{\pi}{2},0,0\right)$ to qubit $0$, $\hat{U}\left(\frac{\pi}{2},0,\pi\right)$ to qubit $1$, $\hat{U}\left(\frac{\pi}{2},0,\frac{\pi}{2}\right)$ to qubit $2$, and $\hat{U}\left(\frac{\pi}{2},0,\frac{\pi}{2}\right)$ to qubit $3$. Figure produced with https://quantum-computing.ibm.com.\label{fig:scheme_circuit_constant}}
\end{figure*}\\
The circuit applied when the function was constant is shown in Fig.~\ref{fig:scheme_circuit_constant}, where no gate was used to create the oracle, as the circuit is constant, mathematically represented by the identity operator, $\hat{I}$. The unitary operators present in Fig.~\ref{fig:scheme_circuit_constant} are used to post-select. They are defined as 
\begin{equation}
\hat{U}\left(\theta,\phi,\lambda\right)=
\begin{pmatrix}
\cos{\frac{\theta}{2}} & -e^{-i\lambda}\sin{\frac{\theta}{2}} \\
e^{i\phi}\sin{\frac{\theta}{2}} & e^{\phi+\lambda}\cos{\frac{\theta}{2}}
\end{pmatrix}.
\end{equation} 
Basically, we associate the post-selected state to the result $\ket{000}$, which implies that finding the system in that state would be equivalent to having post-selected on the state $\ket{\psi_f}$. The circuit applied when the function is balanced, with $f\left(00\right)=f\left(01\right)=0$ and $f\left(10\right)=f\left(11\right)=1$, is shown in Fig.~\ref{fig:scheme_circuit_balanced}. The modular value is equal to $1$ when the function is constant, with $\hat{\sigma}^m_y=0$, and equal to $i$ when the function is balanced, with $\hat{\sigma}^m_y=-1$. The visibility was $1$ for the balanced case and $0$ for the constant case.
\\
Fig.~\ref{fig:results_ibm_computer} and Fig.~\ref{fig:results_ibm_computer_simulator} portray the averages of the outcomes of the $\hat{\sigma}_y$ measurements required to solve the Deutsch-Jozsa problem within the framework of the modular-value algorithm, when executed on the IBM Quito quantum computer (ibmq\_quito) and on the IBM simulator\_mps (ibmq\_simulator\_mps). These results are exhibited as a function of the repetition count. Correspondingly, Fig.~\ref{fig:post-selected_rate} and Fig.~\ref{fig:post-selected_rate_simulator} present the post-selection rates for both the balanced and constant cases run in the quantum computer and in a simulator. It is noteworthy that the theoretical post-selection probability stands at first order at $\frac{1}{2}$, while the value computed through the IBM quantum simulator reaches $0.51$. In the ibmq\_quito, the post-selection rate varies between $0.4$ and $0.6$, while in the simulator (no errors are involved), the post-selected rate fluctuates much less, between $0.47$ and $0.53$. The determination of the requisite number of repetitions is contingent upon the desired data quality and the precision of the experimental setup. Remarkably, the distinction between the two cases—constant and balanced—was successfully accomplished with $32$ shots, incorporating unsuccessful post-selections as well.
A pivotal observation is that the ibmq\_quito system is markedly influenced by substantial noise, as evidenced by a median CNOT error of 6.895e-3 and a median readout error of 3.920e-2. If no errors were involved in the quantum computer, $8$ shots would be enough to differentiate between balanced and constant, as we can see in Fig.~\ref{fig:results_ibm_computer_simulator}. It is worth emphasizing that this procedure can also be conducted utilizing alternative experimental configurations, such as optical systems.\\
\section{Conclusions and perspectives}
We have introduced a novel approach for executing quantum algorithms by using modular values. This technique is applicable to various quantum algorithms that involve an oracle. An additional qubit is needed to read out the modular value result. The number of shots required depends on the post-selection probability and visibility. This procedure can be applied to different experimental setups such as quantum computers or optical systems.\\
This paper presents a method for implementing algorithms by using modular values in quantum computers. Despite the considerable noise in the ibmq\_quito quantum computer, with errors on the order of $10^{-3}$ for quantum gates and $10^{-2}$ for measuring, we have achieved preliminary results for the Deutsch-Jozsa problem by using modular values. No advantages were observed in that particular case. It requires a significantly larger number of shots compared to the standard algorithm, where only one shot is typically sufficient (if no errors are considered). While there are various ways to potentially improve the algorithm, such as selecting appropriate pre- and post-selected states, it is unlikely to yield any significant advantages in this scenario. The probability of post-selection consistently hinders the efficiency of the modular value-based algorithm in reaching the correct result expeditiously.\\
The modular value-based approach demonstrates significant potential as a more precise alternative to various algorithms. For instance, in the Deutsch-Jozsa algorithm, it can discern between the various manners of achieving a balanced function. Moreover, this method could also facilitate the development of new algorithms provided a quantitative validation is performed.\\
In general, we have not yet found any experimental computational advantage over conventional quantum algorithms. Several questions need to be addressed to explore this approach further. First, we need to study all the information provided by the measurements in the computational basis to reduce the number of shots required. Post-selection in the different states of the computational basis simultaneously could be a possible approach, as we directly obtain the information when running an algorithm on a quantum computer. Additionally, we need to investigate the selection of pre- and post-selected states to increase the algorithm's efficiency, specifically by examining the states' separability. As we have seen, non-separable states provide a larger parameter space. However, these states are much more complicated to create by using the common quantum gates. Separable states can be created by using an individual unitary transformation in each qubit. This approach is not always possible for non-separable states. From a computational perspective, we need to improve our ability to create non-separable states by using standard gates.\\
Finally, modular values are expected to yield good results in algorithms that search for small quantities due to their amplification capability. For example, we could differentiate between two almost identical gates. Even though, in that case, the probability of post-selection would be very small, the precision that one can reach by using a modular value-based approach could be better than by using other standard approaches, at least in the presence of technical limitations. In the counterpart, the required number of repetitions of the algorithm could be very large.\\
In summary, exploring the applicability of modular values in various contexts could yield valuable insights and advancements in the field of quantum computing.

\section*{Acknowledgments}
Y.C. is a research associate of the Fund for Scientific Research (F.R.S.-FNRS). This research was supported by the Action de Recherche Concertée WeaM at the University of Namur (19/23-001). We acknowledge the use of IBM Quantum services for this work. The views expressed are those of the authors and do not reflect the official policy or position of IBM or the IBM Quantum team.

\bibliographystyle{unsrt} 
\bibliography{biblio} 

\begin{thebibliography}{10}

\bibitem{grover1996fast}
Lov~K Grover.
\newblock A fast quantum mechanical algorithm for database search.
\newblock In {\em Proceedings of the twenty-eighth annual ACM symposium on
  Theory of computing}, pages 212--219, 1996.

\bibitem{shor1994algorithms}
Peter~W Shor.
\newblock Algorithms for quantum computation: discrete logarithms and
  factoring.
\newblock In {\em Proceedings 35th annual symposium on foundations of computer
  science}, pages 124--134. Ieee, 1994.

\bibitem{deutsch1992rapid}
David Deutsch and Richard Jozsa.
\newblock Rapid solution of problems by quantum computation.
\newblock {\em Proceedings of the Royal Society of London. Series A:
  Mathematical and Physical Sciences}, 439(1907):553--558, 1992.

\bibitem{cleve1998quantum}
Richard Cleve, Artur Ekert, Chiara Macchiavello, and Michele Mosca.
\newblock Quantum algorithms revisited.
\newblock {\em Proceedings of the Royal Society of London. Series A:
  Mathematical, Physical and Engineering Sciences}, 454(1969):339--354, 1998.

\bibitem{aharonov1988result}
Yakir Aharonov, David~Z Albert, and Lev Vaidman.
\newblock How the result of a measurement of a component of the spin of a
  spin-1/2 particle can turn out to be 100.
\newblock {\em Physical review letters}, 60(14):1351, 1988.

\bibitem{kedem2010modular}
Yaron Kedem and Lev Vaidman.
\newblock Modular values and weak values of quantum observables.
\newblock {\em Physical review letters}, 105(23):230401, 2010.

\bibitem{kocsis2011observing}
Sacha Kocsis, Boris Braverman, Sylvain Ravets, Martin~J Stevens, Richard~P
  Mirin, L~Krister Shalm, and Aephraim~M Steinberg.
\newblock Observing the average trajectories of single photons in a two-slit
  interferometer.
\newblock {\em Science}, 332(6034):1170--1173, 2011.

\bibitem{matzkin2019weak}
A~Matzkin.
\newblock Weak values and quantum properties.
\newblock {\em Foundations of Physics}, 49(3):298--316, 2019.

\bibitem{Rafiepoor2021}
E~Rafiepoor, M~R Bazrafkan, and S~Batebi.
\newblock The nonclassicality of a quantum system in weak measurement by means
  of modular values.
\newblock {\em J. Mod. Opt.}, 68(5):295--301, March 2021.

\bibitem{Cormann2017}
Mirko Cormann and Yves Caudano.
\newblock Geometric description of modular and weak values in discrete quantum
  systems using the majorana representation.
\newblock {\em J. Phys. A Math. Theor.}, 50(30):305302, June 2017.

\bibitem{Lundeen2011}
Jeff~S Lundeen, Brandon Sutherland, Aabid Patel, Corey Stewart, and Charles
  Bamber.
\newblock Direct measurement of the quantum wavefunction.
\newblock {\em Nature}, 474(7350):188--191, June 2011.

\bibitem{Pan2019}
Wei-Wei Pan, Xiao-Ye Xu, Yaron Kedem, Qin-Qin Wang, Zhe Chen, Munsif Jan, Kai
  Sun, Jin-Shi Xu, Yong-Jian Han, Chuan-Feng Li, and Guang-Can Guo.
\newblock Direct measurement of a nonlocal entangled quantum state.
\newblock {\em Phys. Rev. Lett.}, 123(15), October 2019.

\bibitem{Turek2020}
Yusuf Turek.
\newblock Direct measurement methods of density matrix of an entangled quantum
  state.
\newblock {\em J. of Phys. Commun.}, 4(7):075007, July 2020.

\bibitem{hosten2008observation}
Onur Hosten and Paul Kwiat.
\newblock Observation of the spin hall effect of light via weak measurements.
\newblock {\em Science}, 319(5864):787--790, 2008.

\bibitem{zhang2015precision}
Lijian Zhang, Animesh Datta, and Ian~A Walmsley.
\newblock Precision metrology using weak measurements.
\newblock {\em Physical review letters}, 114(21):210801, 2015.

\bibitem{li2018chiral}
Dongmei Li, Tian Guan, Yonghong He, Fang Liu, Anping Yang, Qinghua He, Zhiyuan
  Shen, and Meiguo Xin.
\newblock A chiral sensor based on weak measurement for the determination of
  proline enantiomers in diverse measuring circumstances.
\newblock {\em Biosensors and Bioelectronics}, 110:103--109, 2018.

\bibitem{Ho2019}
Le~Bin Ho and Yasushi Kondo.
\newblock Modular-value-based metrology with spin coherent pointers.
\newblock {\em Phys. Lett. A}, 383(2-3):153--157, January 2019.

\bibitem{Ho2016}
Le~Bin Ho and Nobuyuki Imoto.
\newblock Full characterization of modular values for finite-dimensional
  systems.
\newblock {\em Phys. Lett. A}, 380(25-26):2129--2135, June 2016.

\bibitem{Ho2017}
Le~Bin Ho and Nobuyuki Imoto.
\newblock Generalized modular-value-based scheme and its generalized modular
  value.
\newblock {\em Phys. Rev. A}, 95(3), March 2017.

\bibitem{Ho2018}
Le~Bin Ho and Nobuyuki Imoto.
\newblock Various pointer states approaches to polar modular values.
\newblock {\em J. of Math. Phys.}, 59(4), April 2018.

\bibitem{Parks2018}
A~D Parks, S~E Spence, and J~M Farinholt.
\newblock A note concerning the modular valued von neumann interaction
  operator.
\newblock {\em uantum Stud. Math. Found.}, 6(1):101--105, May 2018.

\bibitem{Ogawa2019}
Kazuhisa Ogawa, Osamu Yasuhiko, Hirokazu Kobayashi, Toshihiro Nakanishi, and
  Akihisa Tomita.
\newblock A framework for measuring weak values without weak interactions and
  its diagrammatic representation.
\newblock {\em New J. Phys.}, 21(4):043013, April 2019.

\bibitem{Li2020}
Xiaogang Li and Jiancun Tao.
\newblock Measurement of modular values and connection to the annihilation
  operator.
\newblock {\em {EPL}}, 131(5):50003, September 2020.

\bibitem{pati2019super}
Arun~Kumar Pati.
\newblock Super quantum search algorithm with weak value amplification and
  postselection.
\newblock {\em arXiv preprint arXiv:1910.12390}, 2019.

\bibitem{kashefi2002comparison}
Elham Kashefi, Adrian Kent, Vlatko Vedral, and Konrad Banaszek.
\newblock Comparison of quantum oracles.
\newblock {\em Physical Review A}, 65(5):050304, 2002.

\bibitem{menon2021quantum}
Vikram Menon and Ayan Chattopadhyay.
\newblock Quantum pattern matching oracle construction.
\newblock {\em Pramana}, 95(1):1--3, 2021.

\bibitem{cormann2016revealing}
Mirko Cormann, Mathilde Remy, Branko Kolaric, and Yves Caudano.
\newblock Revealing geometric phases in modular and weak values with a quantum
  eraser.
\newblock {\em Physical Review A}, 93(4):042124, 2016.

\end{thebibliography}
\end{document}